\def\year{2021}\relax
\documentclass[letterpaper]{article} 
\usepackage{aaai21}  
\usepackage{times}  
\usepackage{helvet} 
\usepackage{courier}  
\usepackage[hyphens]{url}  
\usepackage{graphicx} 
\usepackage{subfigure}
\usepackage{algorithm}
\usepackage{algorithmic}
\usepackage{amsfonts}
\usepackage[switch]{lineno}
\usepackage{natbib}  
\usepackage{caption} 
\title{Simplification of Graph Convolutional Networks: \\A Matrix Factorization-based Perspective}

  \author{
      Qiang Liu\textsuperscript{\rm 1,2}, Haoli Zhang\textsuperscript{\rm 1}, Zhaocheng Liu\textsuperscript{\rm 1}\\
  }
  \affiliations{
      \textsuperscript{\rm 1}RealAI        \textsuperscript{\rm 2}Tsinghua University\\
      \{qiang.liu,haoli.zhang,zhaocheng.liu\}@realai.ai
  }

\begin{document}

\maketitle

\begin{abstract}
In recent years, substantial progress has been made on Graph Convolutional Networks (GCNs).
However, the computing of GCN usually requires a large memory space for keeping the entire graph.
In consequence, GCN is not flexible enough, especially for complex real-world applications.
Fortunately, for transductive graph representation learning, methods based on Matrix Factorization (MF) naturally support constructing mini-batches, and thus are more friendly to distributed computing compared with GCN.
Accordingly, in this paper, we analyze the connections between GCN and MF, and simplify GCN as matrix factorization with unitization and co-training.
Furthermore, under the guidance of our analysis, we propose an alternative model to GCN named Unitized and Co-training Matrix Factorization (UCMF).
Extensive experiments have been conducted on several real-world datasets.
On the task of semi-supervised node classification, the experimental results illustrate that UCMF achieves similar or superior performances compared with GCN.
Meanwhile, distributed UCMF significantly outperforms distributed GCN methods, which shows that UCMF can greatly benefit real-world applications.
Moreover, we have also conducted experiments on a typical task of graph embedding, i.e., community detection, and the proposed UCMF model outperforms several representative graph embedding models.
\end{abstract}

\section{Introduction}

Nowadays, works on graph convolutional networks (GCNs) \cite{kipf2016semi} have achieved great success in many graph-based tasks, e.g., semi-supervised node classification \cite{velivckovic2017graph,li2018deeper,wu2019net}, unsupervised graph representation learning \cite{hassani2020contrastive,zhu2020deep}, link prediction \cite{Zhang2018Link}, clustering \cite{bo2020structural} and recommendation \cite{Ying2018Graph,wang2019neural,yu2020TAGNN}.
GCN defines a graph convolution operation, which generates the embedding of each node by aggregating the representations of its neighbors. Given a graph, GCN performs the graph convolution operation layer by layer to obtain the final node representations, which are passed to neural networks to support various tasks.

However, as the computing of GCN requires to store the entire adjacency matrix of a graph \cite{chiang2019cluster}, it is hard to perform GCN on large scale real-world complex graphs, where we usually have a constrained memory size and need distributed computing.
Accordingly, GCN is not flexible enough, and needs to be simplified while retaining the high performance. For example, various sampling methods have been proposed \cite{hamilton2017inductive,chen2018fastgcn,Ying2018Graph} to simplify GCN via reducing the number of edges in the graph.
These methods can be performed in mini-batches, but need to sample high-order neighbours of each node and require a high computational cost that exponentially grows with the number of graph convolution layers, as pointed in \cite{chiang2019cluster}.
Instead of sampling, Cluster-GCN \cite{chiang2019cluster} proposes an approach to convert computation on a huge adjacency matrix to computing on a set of small sub-matrices. However, Cluster-GCN still suffers from performance loss when conducting distributed computing, due to the ignoring of some connections in the graph.
Simple Graph Convolution (SGC) \cite{wu2019simplifying} removes nonlinearities and collapses weight matrices between consecutive layers in GCN, which results in continuous multiplication of adjacency matrices, and we can obtain a final linear model. This simplifies GCN, and makes GCN applicable for distributed computing.
However, as mentioned in previous works \cite{li2018deeper,chen2020measuring}, GCN greatly suffers from over-smoothing. And due to the continuous multiplication of adjacency matrices, SGC may aggravate the degree of over-smoothing.
Accordingly, we need a simplified alternative to GCN, which is flexible enough for distributed computing in real-world applications, and can achieve similar or superior performances compared with the original GCN model.

Besides GCN, for transductive graph representation learning, graph embedding methods \cite{perozzi2014deepwalk,tang2015line,grover2016node2vec} are also widely applied.
In general, these methods aim to embed very large graphs into low-dimensional vector spaces, to preserve the structure of graphs. As for GCN, previous work shows that the graph convolution operation is actually a special form of Laplacian smoothing \cite{li2018deeper}. In this way, as the converging of the GCN model, the smoothing process can keep the final representation of a node more and more similar to those of its neighbors. Therefore, GCN is consistent with graph embedding methods in capturing the structural information. According to previous work \cite{qiu2018network}, graph embedding methods have been successfully unified as Matrix Factorization (MF).
Meanwhile, compared with GCN, MF-based methods are extremely flexible and suitable for large scale distributed computing \cite{gemulla2011large,yu2014parallel}.
These methods are also easy to be extended to various complex applications \cite{rendle2011fast,liu2015cot,wu2016contextual}.
Consequently, if we can simplify the GCN model as a special form of MF, large scale and complex real-world applications will greatly benefit from this.

In this paper, we analyze the connections between GCN and MF, and simplify GCN as matrix factorization with unitization and co-training. Here, the unitization indicates conducting vector unitization on node representations, i.e., forcing the norm of each node representation to one. And the co-training process means co-training with the classification task of labeled nodes, as in some previous works \cite{Weston2012Deep,Yang2016Revisiting}. Then, according to our analysis, we formally propose an alternative model to GCN named Unitized and Co-training Matrix Factorization (UCMF) \footnote{Code avalable at: \url{https://github.com/johnlq/UCMF}.}.

We have conducted extensive experiments on several real-world graphs. The experimental results show that unitization and co-training are two essential components of UCMF. Under centralized computing settings, UCMF achieves similar or superior performances compared with GCN and SGC on the task of semi-supervised node classification. Meanwhile, both GCN and SGC perform poor on graphs that are relatively dense, while UCMF has great performances. This may be caused by the over-smoothing of graph convolution on dense graphs, while UCMF can balance the smoothing of neighbours and the classification of labeled nodes through the co-training process. Experiments under distributed computing settings are also conducted, where UCMF significantly outperforms distributed GCN methods. We have also conducted experiments on the task of community detection, and UCMF achieves better performances compared with several representative graph embedding models.
These results clearly show that, we can use our proposed UCMF model as a simplified alternative to the original GCN model in real-world applications for transductive graph representation learning.

The main contributions of this paper are summarized as follows:
\begin{itemize}
\item We analyze the connections between GCN and MF, and simplify GCN as a special form of matrix factorization with unitization and co-training.
\item We propose an alternative model to GCN, i.e., unitized and co-training matrix factorization.
\item On the task of semi-supervised node classification, extensive experiments have been conducted on several real-world datasets under both centralized and distributed computing settings, and demonstrate the effectiveness and flexibility of UCMF. Meanwhile, on the task of community detection, UCMF outperforms several representative graph embedding models, e.g., LINE \cite{tang2015line} and DeepWalk \cite{perozzi2014deepwalk}.
\end{itemize}

\section{GCN as Unitized and Co-training MF}

In this section, we are going to simplify GCN as a special form of matrix factorization. First, we start from the analysis of how node representations are learned in GCN. Then, we successfully simplify GCN as matrix factorization with unitization and co-training.

\subsection{Graph Convolutional Networks}

According to the definition in \cite{kipf2016semi}, we can formulate each layer of GCN as
\begin{equation} \label{equation:GCN}
	\mathop {\bf{H}}\nolimits^{(l + 1)}  = \sigma \left( {\mathop {\mathop {\bf{D}}\limits^ \sim  }\nolimits^{ - \frac{1}{2}} \mathop {\bf{A}}\limits^ \sim  \mathop {\mathop {\bf{D}}\limits^ \sim  }\nolimits^{ - \frac{1}{2}} \mathop {\bf{H}}\nolimits^{(l)} \mathop {\bf{W}}\nolimits^{(l+1)} } \right),
\end{equation}
where $\mathop {\bf{A}}\limits^ \sim   = {\bf{A}} + \mathop {\bf{I}}\nolimits_N$ is the adjacency matrix of the graph $G$ with added self-connections, $\mathop {\bf{I}}\nolimits_N $ is the identity matrix for $N$ nodes in graph $G$, ${\mathop {\bf{D}}\limits^ \sim  }$ is a diagonal degree matrix with $\mathop {\mathop {\bf{D}}\limits^ \sim  }\nolimits_{i,i}  = \sum\nolimits_j {\mathop {\mathop {\bf{A}}\limits^ \sim  }\nolimits_{i,j} }$, ${\mathop {\bf{H}}\nolimits^{(l)} }$ is the representation of each node at layer $l$, ${\mathop {\bf{W}}\nolimits^{(l+1)} }$ is a layer-specific trainable weight matrix, and $\sigma \left(  \cdot  \right)$ denotes an activation function (such as ${\mathop{\rm ReLU}\nolimits} \left(  \cdot  \right) = \max \left( {0, \cdot } \right)$).
	
For node classification task, we can obtain a classification loss
\begin{equation} \label{equation:class_loss}
	\mathop l\nolimits_{{\rm{c}}}  = {\mathop{\rm CrossEntropy}\nolimits} \left( {{\bf{Y}},\mathop{\rm softmax} \left( {\mathop {\bf{H}}\nolimits^{( - 1)} } \right)} \right),
\end{equation}
where ${\bf{Y}}$ is the ground truth labels for the classification task, ${\mathop {\bf{H}}\nolimits^{(-1)} }$ is the representation of each node at the final layer of GCN. Via optimizing Eq. (\ref{equation:class_loss}), the cross-entropy error of the node classification task can be minimized, and the GCN model can be learned.

\subsection{Simplification} \label{sec:simple}

In \cite{li2018deeper}, GCN has been proved to be a special form of Laplacian smoothing.
As the GCN model goes deeper and we have more layers of graph convolution, the representations in Eq. (\ref{equation:GCN}) have a termination condition as
\begin{equation} \label{equation:termination}
	\mathop {\bf{H}}\nolimits^{(-1)}  = \sigma \left( {\mathop {\mathop {\bf{D}}\limits^ \sim  }\nolimits^{ - \frac{1}{2}} \mathop {\bf{A}}\limits^ \sim  \mathop {\mathop {\bf{D}}\limits^ \sim  }\nolimits^{ - \frac{1}{2}} \mathop {\bf{H}}\nolimits^{(-1)} \mathop {\bf{W}}\nolimits^{(-1)} } \right),
\end{equation}
where $\mathop {\bf{H}}\nolimits^{(-1)}$ is the final representations on the last layer of GCN.
In the simplification discussed in SGC \cite{wu2019simplifying}, nonlinear activation function is ignored.
Meanwhile, according to the implementation in some previous works \cite{kipf2016semi,velivckovic2017graph,wu2019net}, there is no activation function on the last layer of GCN.
Thus, an approximate solution of Eq. (\ref{equation:termination}) can be written as
\begin{equation} \label{equation:converge_graph}
	\mathop {\bf{H}}\nolimits^{( - 1)}  = \mathop {\mathop {\bf{D}}\limits^ \sim  }\nolimits^{ - \frac{1}{2}} \mathop {\bf{A}}\limits^ \sim  \mathop {\mathop {\bf{D}}\limits^ \sim  }\nolimits^{ - \frac{1}{2}} \mathop {\bf{H}}\nolimits^{( - {\rm{1}})}.
\end{equation}
More specifically, for each node $i$ in graph $G$, the approximate solution of the corresponding final representation is
\begin{equation} \label{equation:converge_node_complex}
\small
\mathop h\nolimits_i^{\left( { - 1} \right)}  = \sum\limits_{j \in I} {\frac{1}{{\sqrt {\left( {\mathop d\nolimits_i  + 1} \right)\left( {\mathop d\nolimits_j  + 1} \right)} }}\mathop {\bf{A}}\nolimits_{i,j} \mathop h\nolimits_j^{\left( { - 1} \right)} }  + \frac{1}{{\mathop d\nolimits_i  + 1}}\mathop h\nolimits_i^{\left( { - 1} \right)},
\end{equation}
from which we have
\begin{equation} \label{equation:converge_node}
\mathop h\nolimits_i^{\left( { - 1} \right)}  = \sum\limits_{j \in I} {\frac{1}{{\mathop d\nolimits_i }}\sqrt {\frac{{\mathop d\nolimits_i  + 1}}{{\mathop d\nolimits_j  + 1}}} \mathop {\bf{A}}\nolimits_{i,j} \mathop h\nolimits_j^{\left( { - 1} \right)} },
\end{equation}
where $I$ denotes the set of all the nodes in graph $G$, and ${\mathop d\nolimits_i }$ is the degree of node $i$.

According to above analysis, to train an approximate GCN model,
which can simultaneously model the structure of graph convolution and the node classification task, we can minimize the following loss function
\begin{equation} \label{equation:loss_all}
	l = \alpha \mathop l\nolimits_{{\rm{c}}}  + \left( {1 - \alpha } \right)\mathop l\nolimits_{s},
\end{equation}
where $\alpha$ is a hyper-parameter to control the balance between the two losses, and the structure loss $\mathop l\nolimits_{s}$ refers to
\begin{equation} \label{equation:loss_structure}
\mathop l\nolimits_s {\rm{ = }}\sum\limits_{i \in I} {{\mathop{\rm dis}\nolimits} \left( {\mathop h\nolimits_i^{\left( { - 1} \right)} ,\sum\limits_{j \in I} {\frac{1}{{\mathop d\nolimits_i }}\sqrt {\frac{{\mathop d\nolimits_i  + 1}}{{\mathop d\nolimits_j  + 1}}} \mathop {\bf{A}}\nolimits_{i,j} \mathop h\nolimits_j^{\left( { - 1} \right)} } } \right)},
\end{equation}
where ${\mathop{\rm dis}\nolimits} \left( { \cdot , \cdot } \right)$ is a distance measurement. Here, we adopt the commonly-used cosine distance, and obtain
\begin{equation}
\mathop l\nolimits_s {\rm{ = }}\sum\limits_{i \in I} {{\mathop{\rm Cosine}\nolimits} \left( {\mathop h\nolimits_i^{\left( { - 1} \right)} ,\sum\limits_{j \in I} {\frac{1}{{\mathop d\nolimits_i }}\sqrt {\frac{{\mathop d\nolimits_i  + 1}}{{\mathop d\nolimits_j  + 1}}} \mathop {\bf{A}}\nolimits_{i,j} \mathop h\nolimits_j^{\left( { - 1} \right)} } } \right)},
\end{equation}
which is equivalent to
\begin{equation} \label{equation:loss_structure_cos_1}
\mathop l\nolimits_s  =  - \sum\limits_{i \in I} {\frac{{\mathop h\nolimits_i^{\left( { - 1} \right)} \mathop {\left( {\sum\limits_{j \in I} {\frac{1}{{\mathop d\nolimits_i }}\sqrt {\frac{{\mathop d\nolimits_i  + 1}}{{\mathop d\nolimits_j  + 1}}} \mathop {\bf{A}}\nolimits_{i,j} \mathop h\nolimits_j^{\left( { - 1} \right)} } } \right)}\nolimits^ \top  }}{{\left\| {\mathop h\nolimits_i^{\left( { - 1} \right)} } \right\|\left\| {\sum\limits_{j \in I} {\frac{1}{{\mathop d\nolimits_i }}\sqrt {\frac{{\mathop d\nolimits_i  + 1}}{{\mathop d\nolimits_j  + 1}}} \mathop {\bf{A}}\nolimits_{i,j} \mathop h\nolimits_j^{\left( { - 1} \right)} } } \right\|}}}.
\end{equation}

\begin{figure}
	\centering
	\includegraphics[width=0.48\textwidth]{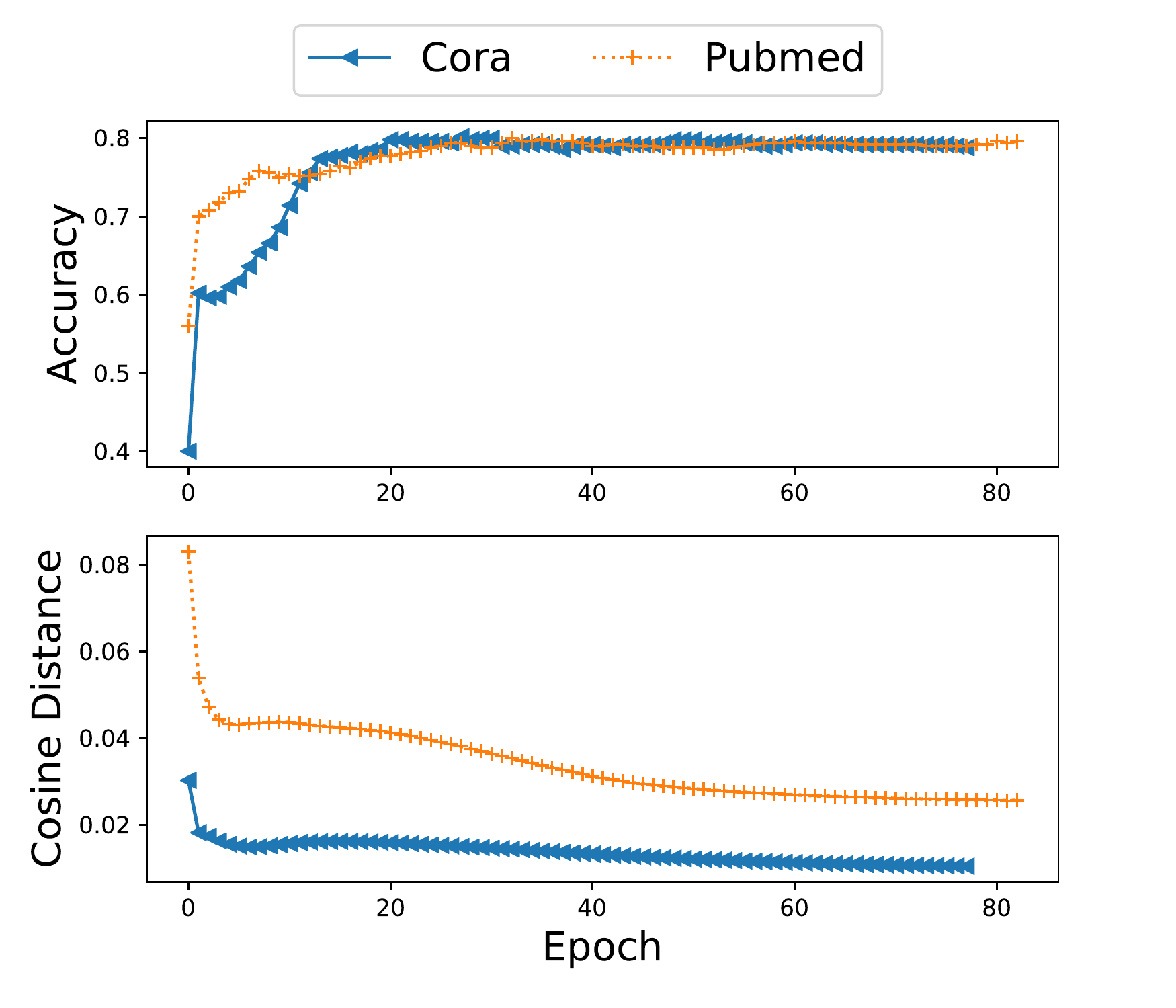}
	\caption{Consistency estimation between the changing of cosine distance and the convergence of GCN during the training procedure on the Cora dataset and the Pubmed dataset. The upper part shows the accuracy curves during the training of GCN. The lower part shows the average cosine distance between representations of nodes in the graph and those of their neighbors during the training of GCN.}
	\label{fig:cos}
\end{figure}

To verify whether the changing of cosine distance is consistent with the convergence of GCN during the training procedure, we conduct empirical experiments and train GCN models on the Cora dataset and the Pubmed dataset. Fig. \ref{fig:cos} demonstrates the average cosine distance between representations of nodes in the graph and those of their neighbors, as well as the convergence curves estimated by accuracy during the training of GCN on the two datasets. It is obvious that, the tendencies of curves on the same dataset match with each other. That is to say, the changing of cosine distance is consistent with the convergence of the GCN model.

Then, to simplify the form of Eq. (\ref{equation:loss_structure_cos_1}), we conduct vector unitization on the learned representations $\mathop {\bf{H}}\nolimits^{( - 1)}$, and thus each representation $\mathop h\nolimits_i^{\left( { - 1} \right)}$ has similar l2-norm.
As a result, through unitization, Eq. (\ref{equation:loss_structure_cos_1}) is equivalent to
\begin{equation} \label{equation:loss_WMF}
	\mathop l\nolimits_{s}  =  - \sum\limits_{i \in I} {\sum\limits_{j \in \mathop C\nolimits_i } {\frac{1}{{\mathop d\nolimits_i }}\sqrt {\frac{{\mathop d\nolimits_i  + 1}}{{\mathop d\nolimits_j  + 1}}} \mathop {\bf{A}}\nolimits_{i,j} \mathop v\nolimits_i \mathop v\nolimits_j^ \top  } },
\end{equation}
where ${\mathop C\nolimits_i }$ denotes all the nodes that node $i$ is connected to, and $\mathop v\nolimits_i  = \mathop h\nolimits_i^{\left( { - 1} \right)}$ for simplicity. Moreover, for better optimization, we can incorporate negative log likelihood and minimize the following loss function equivalently to Eq. (\ref{equation:loss_WMF})
\begin{equation} \label{equation:loss_WMF_1}
	\mathop l\nolimits_{s}  =  - \sum\limits_{i \in I} {\sum\limits_{j \in \mathop C\nolimits_i } {\frac{1}{{\mathop d\nolimits_i }}\sqrt {\frac{{\mathop d\nolimits_i  + 1}}{{\mathop d\nolimits_j  + 1}}} \mathop {\bf{A}}\nolimits_{i,j} \log \left( {\lambda \left( {\mathop v\nolimits_i \mathop v\nolimits_j^ \top  } \right)} \right)} },
\end{equation}
where $\lambda \left(  \cdot  \right) = {\mathop{\rm sigmoid}\nolimits} \left(  \cdot  \right)$.

Usually, in graph embedding methods \cite{perozzi2014deepwalk,tang2015line,grover2016node2vec}, negative sampling of edges is used, for better convergence. Thus, we can randomly sample negative edges for each edge in graph $G$.
Following previous works in unifying word embedding \cite{levy2014neural} and graph embedding \cite{qiu2018network} as implicit matrix factorization, we can rewrite Eq.~(\ref{equation:loss_WMF_1}) as
\begin{equation} \label{equation:loss_WMF_3}
\begin{array}{l}
\mathop l\nolimits_{s}  =  - \sum\limits_{i \in I} {\sum\limits_{j \in \mathop C\nolimits_i } {\mathop \beta \nolimits_{i,j} \mathop {\bf{A}}\nolimits_{i,j} \log \left( {\lambda \left( {\mathop v\nolimits_i \mathop v\nolimits_j^ \top  } \right)} \right)} } \\
\quad \quad - \sum\limits_{i \in I} {k d_i \mathbb E_{j' \sim \mathop P\nolimits_G } \left[ {\mathop \beta \nolimits_{i,j'} \log \left( {\lambda \left( { - \mathop v\nolimits_i \mathop v\nolimits_{j'}^ \top  } \right)} \right)} \right]}
\end{array},
\end{equation}
where $\mathop \beta \nolimits_{i,j}  = \mathop d\nolimits_i^{ - 1} \mathop {\left( {\mathop d\nolimits_i  + 1} \right)}\nolimits^{1/2} \mathop {\left( {\mathop d\nolimits_j  + 1} \right)}\nolimits^{ - 1/2}$, $k$ is the number of negative samples for each edge, and ${\mathop P\nolimits_G }$ denotes the distribution that generates negative samples in graph $G$. For each node $i$, $\mathop P\nolimits_G \left( i \right) = \mathop d\nolimits_i /\left| G \right|$, where $\left| G \right|$ is the number of edges in graph $G$.
We can explicitly express the expectation term as
\begin{equation}
\scriptsize
\mathbb E_{j' \sim \mathop P\nolimits_G } \left[ {\mathop \beta \nolimits_{i,j'} \log \left( {\lambda \left( { - \mathop v\nolimits_i \mathop v\nolimits_{j'}^ \top  } \right)} \right)} \right] = \sum\limits_{j' \in I} {\frac{{\mathop \beta \nolimits_{i,j'} \mathop d\nolimits_{j'} }}{{\left| G \right|}}\log \left( {\lambda \left( { - \mathop v\nolimits_i \mathop v\nolimits_{j'}^ \top  } \right)} \right)},
\end{equation}
from which we have
\begin{equation} \label{equation:expectation}
\small
\begin{array}{l}
\mathbb E_{j' \sim \mathop P\nolimits_G } \left[ {\mathop \beta \nolimits_{i,j'} \log \left( {\lambda \left( { - \mathop v\nolimits_i \mathop v\nolimits_{j'}^ \top  } \right)} \right)} \right] = \frac{{\mathop \beta \nolimits_{i,j} \mathop d\nolimits_j }}{{\left| G \right|}}\log \left( {\lambda \left( { - \mathop v\nolimits_i \mathop v\nolimits_j^ \top  } \right)} \right)\\
\quad \quad \quad \quad \quad \quad \quad \quad \quad \quad + \sum\limits_{j' \in I\backslash \left\{ j \right\}} {\frac{{\mathop \beta \nolimits_{i,j'} \mathop d\nolimits_{j'} }}{{\left| G \right|}}\log \left( {\lambda \left( { - \mathop v\nolimits_i \mathop v\nolimits_{j'}^ \top  } \right)} \right)}
\end{array}.
\end{equation}
Then, we can obtain the local structure loss for a specific edge $\left( {i,j} \right)$ as
\begin{equation} \label{equation:loss_pair_1}
\begin{array}{l}
\mathop l\nolimits_{s} \left( {i,j} \right) =  - \mathop \beta \nolimits_{i,j} \mathop {\bf{A}}\nolimits_{i,j} \log \left( {\lambda \left( {\mathop v\nolimits_i \mathop v\nolimits_j^ \top  } \right)} \right)\\
\quad \quad \quad \quad- \frac{{k\mathop \beta \nolimits_{i,j} \mathop d\nolimits_i \mathop d\nolimits_j }}{{\left| G \right|}}\log \left( {\lambda \left( { - \mathop v\nolimits_i \mathop v\nolimits_j^ \top  } \right)} \right)
\end{array}.
\end{equation}
To optimize above objective, we need to calculate the partial derivative of $\mathop l\nolimits_{s} \left( i,j \right)$ with respect to ${\mathop v\nolimits_i \mathop v_j^ \top }$
\begin{equation} \label{equation:partial_derivative}
\scriptsize
	\frac{{\partial \mathop l\nolimits_{s} \left( {i,j} \right) }}{{\partial \left( {\mathop v\nolimits_i \mathop v\nolimits_j^ \top  } \right)}} =  - \mathop \beta \nolimits_{i,j} \mathop {\bf{A}}\nolimits_{i,j} \lambda \left( { - \mathop v\nolimits_i \mathop v\nolimits_j^ \top  } \right) + \frac{{k\mathop \beta \nolimits_{i,j} \mathop d\nolimits_i \mathop d\nolimits_j }}{{\left| G \right|}}\lambda \left( {\mathop v\nolimits_i \mathop v\nolimits_j^ \top  } \right).
\end{equation}
Via setting Eq. (\ref{equation:partial_derivative}) to zero, we can obtain
\begin{equation} \label{equation:zero}
	\mathop e\nolimits^{2\mathop v\nolimits_i \mathop v\nolimits_j^ \top  }  - \left( {\frac{{\left| G \right|\mathop {\bf{A}}\nolimits_{i,j} }}{{k\mathop d\nolimits_i \mathop d\nolimits_j }} - 1} \right)\mathop e\nolimits^{\mathop v\nolimits_i \mathop v\nolimits_j^ \top  }  - \frac{{\left| G \right|\mathop {\bf{A}}\nolimits_{i,j} }}{{k\mathop d\nolimits_i \mathop d\nolimits_j }} = 0,
\end{equation}
which has two solutions, $\mathop e\nolimits^{\mathop v\nolimits_i \mathop v\nolimits_j^ \top  }  =  - 1$ and
\begin{equation} \label{equation:solution_final}
	\mathop v\nolimits_i \mathop v\nolimits_j^ \top   = \log \left( {\frac{{\left| G \right|\mathop {\bf{A}}\nolimits_{i,j} }}{{k\mathop d\nolimits_i \mathop d\nolimits_j }}} \right).
\end{equation}

Accordingly, the GCN model can be simplified as the following matrix factorization
\begin{equation} \label{equation:MF}
{\bf{V}} {\bf{V}} ^\top = \log \left( {\left| G \right| {\bf{D}} ^{ - 1} {\bf{A}} {\bf{D}} ^{ - 1} } \right) - \log \left( k \right),
\end{equation}
co-trained with the classification loss $\mathop l\nolimits_{{\rm{c}}}$, where node representations in $\bf{V}$ are unitized. ${\bf{D}}$ is a diagonal degree matrix with ${\bf{D}}_{i,i} = d_i$. According to previous analysis \cite{levy2014neural,qiu2018network,du2018dynamic}, the matrix factorization in Eq. (\ref{equation:MF}) is as the same as common implicit matrix factorization. In summary, we successfully simplify GCN as matrix factorization with unitization and co-training.

\section{Discussion}

In this section, we perform more discussion about our simplification of GCN.

\subsection{Adopting Euler Distance} \label{app:euler}

Besides cosine distance, Euler distance is another commonly-used distance measurement which can be adopted in Eq. (\ref{equation:loss_structure}).
Here, we need to investigate whether the conclusion in Eq. (\ref{equation:MF}) stays the same when Euler distance is adopted.

Suppose we have two node representations $p = \left[ { p_1 ,..., p_d } \right]$ and $q = \left[ { q_1 ,..., q_d } \right]$.
As discussed in the simplification section, we conduct vector unitization on the node representations, which means
\begin{equation} \label{eq:pq}
\sum\limits_{i = 1}^d {\mathop p\nolimits_i^2 }  = 1, ~~~~ \sum\limits_{i = 1}^d {\mathop q\nolimits_i^2 }  = 1.
\end{equation}
The cosine distance between $p$ and $q$ can be formulated as
\begin{equation}
{\rm{Cosine}}\left( {p,q} \right) = 1 - \frac{{p q^ \top  }}{{\left\| p \right\|\left\| q \right\|}}.
\end{equation}
Considering Eq. (\ref{eq:pq}), we have
\begin{equation} \label{eq:cosine}
{\rm{Cosine}}\left( {p,q} \right) = 1 - p q^ \top   = 1 - \sum\limits_{i = 1}^d { p_i q_i }.
\end{equation}
Meanwhile, the Euler distance between $p$ and $q$ can be formulated as
\begin{equation}
\begin{array}{l}
{\rm{Euler}}\left( {p,q} \right) = \sqrt {\sum\limits_{i = 1}^d {\mathop {\left( {\mathop p\nolimits_i  - \mathop q\nolimits_i } \right)}\nolimits^2 } } \\
\quad\quad\quad\quad\quad= \sqrt {\sum\limits_{i = 1}^d {\mathop p\nolimits_i^2 }  + \sum\limits_{i = 1}^d {\mathop q\nolimits_i^2 }  - 2\sum\limits_{i = 1}^d {\mathop p\nolimits_i \mathop q\nolimits_i } }
\end{array}.
\end{equation}
Considering Eq. (\ref{eq:pq}), we can obtain
\begin{equation} \label{eq:euler}
{\rm{Euler}}\left( {p,q} \right) = \sqrt {2 - 2\sum\limits_{i = 1}^d { p_i q_i } }.
\end{equation}
Combining Eq. (\ref{eq:cosine}) and Eq. (\ref{eq:euler}), we can conclude the connection between cosine distance and Euler distance as
\begin{equation}
{\rm{Euler}}\left( {p,q} \right) = \sqrt { 2 {\rm{Cosine}}\left( {p,q} \right)}.
\end{equation}

Accordingly, adopting the Euler distance, the loss function in Eq. (\ref{equation:loss_WMF}) can be rewritten as
\begin{equation} \label{eq:euler_loss}
\mathop l\nolimits_s  = \sum\limits_{i \in I} {\sqrt {2 - 2\mathop v\nolimits_i \mathop {\left( {\sum\limits_{j \in \mathop C\nolimits_i } {\mathop \beta \nolimits_{i,j} \mathop {\bf{A}}\nolimits_{i,j} \mathop v\nolimits_j } } \right)}\nolimits^ \top  } }.
\end{equation}
To optimize Eq. (\ref{eq:euler_loss}), we can optimize the following loss function equivalently
\begin{equation} \label{eq:euler_loss1}
\begin{array}{l}
\mathop l\nolimits_s  = \sum\limits_{i \in I} {\left( {2 - 2\mathop v\nolimits_i \mathop {\left( {\sum\limits_{j \in \mathop C\nolimits_i } {\mathop \beta \nolimits_{i,j} \mathop {\bf{A}}\nolimits_{i,j} \mathop v\nolimits_j } } \right)}\nolimits^ \top  } \right)} \\
\quad= 2\left| V \right| - 2\sum\limits_{i \in I} {\sum\limits_{j \in \mathop C\nolimits_i } {\mathop \beta \nolimits_{i,j} \mathop {\bf{A}}\nolimits_{i,j} \mathop v\nolimits_i \mathop v\nolimits_i^ \top  } }
\end{array},
\end{equation}
where $\left| V \right|$ is the number of nodes in graph $G$.
It is obvious that, the loss function in Eq. (\ref{eq:euler_loss1}) is equivalent to Eq. (\ref{equation:loss_WMF}).
Thus, when we adopt the Euler distance in Eq. (\ref{equation:loss_structure}), we can equivalently obtain the same matrix factorization as in Eq. (\ref{equation:MF}).

\subsection{Another Form of Graph Convolution} \label{app:another}

Besides Eq. (\ref{equation:GCN}), there is another form of graph convolution for GCN \cite{kipf2016semi,li2018deeper}, which can be formulated as
\begin{equation}
	\mathop {\bf{H}}\nolimits^{(l + 1)}  = \sigma \left( {\mathop {\mathop {\bf{D}}\limits^ \sim  }\nolimits^{ -1} \mathop {\bf{A}}\limits^ \sim  \mathop {\bf{H}}\nolimits^{(l)} \mathop {\bf{W}}\nolimits^{(l)} } \right).
\end{equation}
Then, Eq. (\ref{equation:converge_node_complex}) can be rewritten as
\begin{equation}
\mathop h\nolimits_i^{\left( { - 1} \right)}  = \sum\limits_{j \in I} {\frac{1}{{\mathop d\nolimits_i  + 1}}\mathop {\bf{A}}\nolimits_{i,j} \mathop h\nolimits_j^{\left( { - 1} \right)} }  + \frac{1}{{\mathop d\nolimits_i  + 1}}\mathop h\nolimits_i^{\left( { - 1} \right)},
\end{equation}
from which we have
\begin{equation}
\mathop h\nolimits_i^{\left( { - 1} \right)}  = \sum\limits_{j \in I} {\frac{1}{{\mathop d\nolimits_i }}\mathop {\bf{A}}\nolimits_{i,j} \mathop h\nolimits_j^{\left( { - 1} \right)} }.
\end{equation}
Then, Eq. (\ref{equation:loss_WMF}) can be rewritten as
\begin{equation} \label{equation:new_loss_WMF}
	\mathop l\nolimits_{s}  =  - \sum\limits_{i \in I} {\sum\limits_{j \in \mathop C\nolimits_i } {\frac{1}{{\mathop d\nolimits_i }} \mathop {\bf{A}}\nolimits_{i,j} \mathop v\nolimits_i \mathop v\nolimits_j^ \top  } }.
\end{equation}
Accordingly, the only difference between Eq. (\ref{equation:loss_WMF}) and Eq. (\ref{equation:new_loss_WMF}) is that, instead of $\mathop \beta \nolimits_{i,j}  = \mathop d\nolimits_i^{ - 1} \mathop {\left( {\mathop d\nolimits_i  + 1} \right)}\nolimits^{1/2} \mathop {\left( {\mathop d\nolimits_j  + 1} \right)}\nolimits^{ - 1/2}$ in Eq. (\ref{equation:loss_WMF}), we have $\mathop \beta \nolimits_{i,j}  = \mathop d\nolimits_i^{ - 1} $ in Eq. (\ref{equation:new_loss_WMF}).
Meanwhile, $\mathop \beta \nolimits_{i,j} $ can be eliminated during the analysis in Eq. (\ref{equation:loss_WMF_1}-\ref{equation:solution_final}).
It is obvious that, with the new form of graph convolution, we can still obtain the same matrix factorization as in Eq. (\ref{equation:MF}).

\section{The UCMF Architecture} \label{sec:UCMF}

In this section, we formally propose the UCMF architecture.
We first need to deal with node features, which can not be directly handled in the original implicit matrix factorization.
Let $x_i$ denote the feature vector of node $i$, and $f_1  \left(  \cdot  \right)$ denote the first Multi-Layer Perception (MLP) for feature modeling.
According to our analysis, given $x_i$ and $f_1  \left(  \cdot  \right)$, we conduct vector unitization to obtain $v_i$, i.e., the representation of node $i$, as
\begin{equation} \label{equation:sec4_unitization}
	v_i = \frac{f_1(x_i)}{\|f_1(x_i)\|}.
\end{equation}
Then, following our previous analysis, UCMF consists of two losses: the structure loss $\mathop{l}\nolimits_{s}$ and the classification loss $\mathop{l}\nolimits_{c}$. The structure loss $\mathop{l}\nolimits_{s}$ can be formulated as implicit matrix factorization with $k$ negative samples for each edge
\begin{equation} \label{equation:sec4_structure_loss}
\begin{array}{l}
\mathop l\nolimits_{s}  =  - \sum\limits_{i \in I} {\sum\limits_{j \in \mathop C\nolimits_i } { \log \left( {\lambda \left( {\mathop v\nolimits_i \mathop v\nolimits_j^ \top  } \right)} \right)} } \\
 \quad \quad- \sum\limits_{i \in I} {k d_i \mathbb E_{j' \sim \mathop P\nolimits_G } \left[ {\log \left( {\lambda \left( { - \mathop v\nolimits_i \mathop v\nolimits_{j'}^ \top  } \right)} \right)} \right]}
\end{array}.
\end{equation}
Meanwhile, the prediction on node classification can be made as
\begin{equation}
\mathop {\hat y}\nolimits_i  = {\rm{softmax}}\left( {\mathop f\nolimits_2 \left( {\mathop v\nolimits_i } \right)} \right),
\end{equation}
where $f_2  \left(  \cdot  \right)$ is the second MLP for making predictions.
As in GCN, the classification loss $\mathop{l}\nolimits_{c}$ can be obtained as
\begin{equation} \label{equation:sec4_classification_loss}
\mathop l\nolimits_{{\rm{c}}}  = \sum\limits_{i \in \mathop I\nolimits_L } {{\mathop{\rm CrossEntropy}\nolimits} \left( { y_i , {\hat y}_i} \right)},
\end{equation}
where $I_L$ is the set of labeled nodes in the graph, and $y_i$ is the ground-truth label of node $i$. Co-training the two losses as in Eq. (\ref{equation:loss_all}), we obtain the final loss function of the proposed UCMF model.

Furthermore, following some previous works on semi-supervised node classification \cite{Weston2012Deep,Yang2016Revisiting}, during co-training, the two losses $\mathop{l}\nolimits_{s}$ and $\mathop{l}\nolimits_{c}$ are alternately optimized. To be more specific, we first optimize the structure loss $\mathop{l}\nolimits_{s}$ with $b$ batches of samples, then we optimize the classification loss $\mathop{l}\nolimits_{c}$ with one batch of samples. We repeat this process until convergence. Here, the parameter $b$ is the balance parameter between the two losses.

Compared with previous graph modeling methods \cite{Weston2012Deep,Yang2016Revisiting,kipf2016semi,perozzi2014deepwalk,tang2015line,grover2016node2vec}, the most unique part of UCMF is the unitization of node representations, which is derived from our above analysis.
This has also been incorporated in \cite{hamilton2017inductive}.
With unitization, it reduces the possibility of extreme values, and has chance to generate better node representations.

\section{Related Works}

In recent years, GCN \cite{bruna2013spectral,kipf2016semi} has drawn tremendous attention from academia.
It updates node representations with the aggregation of its neighbors. Based on GCN, Graph Attention Network (GAT) \cite{velivckovic2017graph} introduces the attention mechanism to model different influences of neighbors with learnable parameters.

As mentioned in previous works \cite{li2018deeper}, GCN greatly suffers from over-smoothing, and thus it is hard for GCN to go deeper.
In \cite{li2018deeper}, the authors propose to add more supervision for training a deeper GCN.
JK-Nets \cite{xu2018representation} presents general layer aggregation mechanisms to combine the output representation in every GCN layer.
ResGCN and DenseGCN \cite{li2019deepgcns} incorporate some technologies from computer vision, i.e., residual connections \cite{he2016deep} and dense connections \cite{huang2017densely}, to tackle with this problem.
Hierarchical Graph Convolutional Network (H-GCN) \cite{hu2019hierarchical} applies coarsening and refining operations to make GCN deeper.
In \cite{Chen2020Simple}, the authors add initial residual and identity mapping into the GCN model.
In \cite{chen2020measuring}, the authors design some metrics for measuring the degree of over-smoothing, and accordingly propose an approach based on regularization to overcome the problem.

Another severe problem of GCN is that, it requires the entire adjacency matrix of the graph, which makes GCN not flexible on large scale graphs. For sake of flexibility, some works try to simplify GCN from different perspectives.
GraphSAGE \cite{hamilton2017inductive} samples a fixed number
of neighbors for each node in the graph.
FastGCN \cite{chen2018fastgcn} proposes to apply importance sampling to reduce the computation of aggregation of neighbors on graph.
Instead of approximating the node representations, variance controlled GCN \cite{chen2018stochastic} uses sampled node to estimate the change of node representations in every updating step.
These sampling-based methods can be implemented in mini-batches, via sampling high-order neighbours of each node in the graph.
However, as pointed in \cite{chiang2019cluster}, this causes high computational cost growing exponentially with the number of layers.
To tackle with this, Cluster-GCN \cite{chiang2019cluster} uses graph partition method \cite{karypis1998fast} to split the whole graph into a set of small sub-graphs, where neighbour aggregation happens within each small sub-graph.
With this improvement, Cluster-GCN supports constructing mini-batches and distributed computing.
Simple Graph Convolution (SGC) \cite{wu2019simplifying} simplifies GCN to continuous multiplication of adjacency matrices with a linear classifier. Considering SGC is actually a linear model, it is capable to be distributed implemented. However, due to the continuous multiplication of adjacency matrices, SGC may aggravates the degree of over-smoothing.

\section{Experiments}
In this section, we empirically evaluate the performance of the proposed UCMF model. We first describe the datasets and settings of the experiments, then report and analyze the experimental results. Thorough evaluations are conducted to answer the following research questions:
\begin{itemize}
	\item \textbf{RQ1} What are the roles of different components in the UCMF model, i.e., unitization and co-training?
	\item \textbf{RQ2} How are the performances of our UCMF model compared with those of GCN on different datasets?
	\item \textbf{RQ3} How do the two hyper-parameters, i.e., the negative sampling number $k$ and the balance parameter $b$, affect the performances of UCMF?
	\item \textbf{RQ4} Compared with GCN, is our proposed UCMF model more friendly to distributed computing?
	\item \textbf{RQ5} How does UCMF perform on community detection, a typical task of graph embedding?
\end{itemize}


\subsection{Experimental Datasets}
We evaluate our proposed model on five real-world datasets, i.e., Cora, Citeseer, Pubmed, BlogCatalog and Flickr.
Cora, Citeseer and Pubmed \cite{sen2008collective} are three standard citation network benchmark datasets\footnote{\url{https://github.com/tkipf/gcn}}, which are widely used in previous works \cite{kipf2016semi,velivckovic2017graph,chiang2019cluster}.
BlogCatalog and Flickr \cite{huang2017label} are two social network datasets\footnote{\url{https://github.com/xhuang31/LANE}}.
The posted keywords or tags in BlogCatalog and Flickr networks are used as node features.
For the splitting of Cora, Citeseer and Pubmed, we follow the classical settings in previous works \cite{Yang2016Revisiting,kipf2016semi}.
And the splitting of BlogCatalog and Flickr is the same as in \cite{wu2019net}.
Specifically, on BlogCatalog and Flickr, we randomly select 10\% and 20\% of the nodes for training and validation respectively, and the rest 70\% as our testing set.
The sparsity of Cora, Citeseer, Pubmed, BlogCatalog and Flickr is $99.85\%$, $99.91\%$, $99.97\%$, $98.73\%$ and $99.15\%$ respectively.
Accordingly, we have two different categories of datasets: sparse and dense.
Sparse datasets consist of Cora, Citeseer and Pubmed.
Dense datasets consist of BlogCatalog and Flickr.

\subsection{Experimental Settings}

In our experiments, we run each model 10 times with random weight initialization, and report the average evaluation values, as well as statistically significant improvement measured by t-test with p-value $<0.05$.
When we implement UCMF and its extended variations, we set batch size as $256$, the dimensionality of node representation $v_i$ as $10\%$ of the dimensionality of original node features on each dataset, the dropout rate as $0.5$, the l2 regularization as $0.002$, and tune the learning rate in $[0.001, 0.005, 0.01]$.
The first MLP $f_1  \left(  \cdot  \right)$ for feature modeling is with one layer, which outputs the node representations.
And the second MLP $f_2  \left(  \cdot  \right)$ for making predictions is with two layers, where the number of hidden neurons is set as $128$.
Moreover, to answer the five research questions, we have conducted extensive experiments from different perspectives.
We introduce the compared models and detailed settings as follow.

Firstly, we need to investigate the effects of the two components in UCMF, i.e., co-training and unitization.
We conduct experiments on semi-supervised node classification, and report the results in terms of accuracy.
To clarify the contribution of the two components, comparisons are conducted without utilizing node features.
In this experiment, besides the \textbf{UCMF} model, we include another two variations of UCMF into the comparison: \textbf{UCMF-C} and \textbf{UCMF-U}.
UCMF-C means the UCMF model without co-training with the classification loss, and UCMF-U indicates the UCMF model without vector unitization on node representations.
We also include three commonly-used graph embedding methods in our experiment: \textbf{MF}, \textbf{DeepWalk} \cite{perozzi2014deepwalk} and \textbf{node2vec} \cite{grover2016node2vec}, which are incapable to directly take node features into consideration.
The \textbf{GCN} model \cite{kipf2016semi} is also taken into account, where we have two layers of graph convolution.
Considering we do not utilize node features in this comparison, the input node features in GCN and UCMF are replaced with learnable embeddings of each node.

Secondly, to verify the effectiveness of UCMF, we conduct experiments on semi-supervised node classification utilizing node features.
Besides \textbf{UCMF} and \textbf{GCN} \cite{kipf2016semi}, we involve \textbf{Planetoid} \cite{Yang2016Revisiting} as a baseline.
And two simplified GCN models, i.e., \textbf{fastGCN} \cite{chen2018fastgcn} and \textbf{SGC} \cite{wu2019simplifying}, are also compared.
In our experiments, all the GCN-based models, i.e., GCN, fastGCN and SGC, are with two layers of graph convolution.
And we follow the default hyper-parameter settings in the corresponding original papers.

Thirdly, we investigate the effects of the two important hyper-parameters, i.e., the negative sampling number $k$ and the balance parameter $b$, on model performances. In this experiment, node features are utilized. We report the accuracy of \textbf{UCMF} on semi-supervised node classification with respect to different $k$ and $b$.

\begin{table}[t]
\centering
\scriptsize
\begin{tabular}{c|ccccc}
\hline
compared model & Cora  & Citeseer & Pubmed & BlogCatalog & Flickr \\
\hline
MF    & 50.8  & 33.3  & 43.6  & 45.8  & 24.6 \\
DeepWalk & 67.2  & 43.2  & 65.3  & 61.8  & 41.5 \\
node2vec & 67.8  & 43.5  & \textbf{65.8} & 63.1  & 42.2 \\
UCMF-U & 51.2  & 35.4  & 44.4  & 60.9  & 38.6 \\
UCMF-C & 55.1  & 35.5  & 44.2  & 53.1  & 25.1 \\
UCMF  & \textbf{69.1}$^*$ & \textbf{46.2}$^*$ & \textbf{66.2}$^*$ & \textbf{66.2}$^*$ & \textbf{43.2}$^*$ \\
GCN   & 65.6  & 44.4  & 58.3  & 58.3  & 30.9 \\
\hline
\end{tabular}%
\caption{Performance comparison on semi-supervised node classification in terms of classification accuracy (\%) without utilizing node features. The larger the values, the better the performances. Competitive performances on each dataset are highlighted. $*$ denotes statistically significant improvement measured by t-test with p-value $<0.05$.}
\label{tab:comparison1}
\end{table}

Then, to test the performances under distributed computing settings, we compare \textbf{UCMF} with several distributed GCN models on semi-supervised node classification.
In the comparison, we involve the state-of-the-art distributed GCN model \textbf{Cluster-GCN} \cite{chiang2019cluster}, as well as the baseline model \textbf{Random-GCN} in the corresponding paper.
Moreover, \textbf{SGC} \cite{wu2019simplifying} is also involved in the comparison, for it is actually a linear model and can be distributed implemented.
We also run \textbf{fastGCN} \cite{chen2018fastgcn} in mini-batches via sampling high-order neighbours of each node in the graph, though this causes exponential complexity with the number of layers as pointed in \cite{chiang2019cluster}.
Under distributed computing settings, parameters of each model are learned with the Parameter Server (PS) architecture \cite{li2014scaling}, where we have $1$ server and $2$ workers.
Each graph is partitioned into two sub-graphs for computing on the two workers.
For Cluster-GCN, following \cite{chiang2019cluster}, the partition is conducted with the METIS algorithm \cite{karypis1998fast}. And for Random-GCN, fastGCN, SGC and UCMF, the partition is randomly conducted.

Finally, we want to see if UCMF can perform well on typical tasks of graph embedding.
Thus, we conduct experiments on community detection, and report the results in terms of conductance.
Conductance is basically the ratio between the number of edges leaving a community and that within the community.
We perform k-means clustering on node representations to generate communities, where each cluster refers to one community.
The number of communities is the same as the number of classes on each dataset.
We involve several representative graph embedding methods: \textbf{MF}, \textbf{LINE} \cite{tang2015line}, \textbf{DeepWalk} \cite{perozzi2014deepwalk} and \textbf{node2vec} \cite{grover2016node2vec}.
Considering community detection is an unsupervised task and we should not use supervision information, we perform \textbf{UCMF-C} in this experiment.

\subsection{Performance Analysis}
\begin{figure}
\centering
\includegraphics[width=0.48\textwidth]{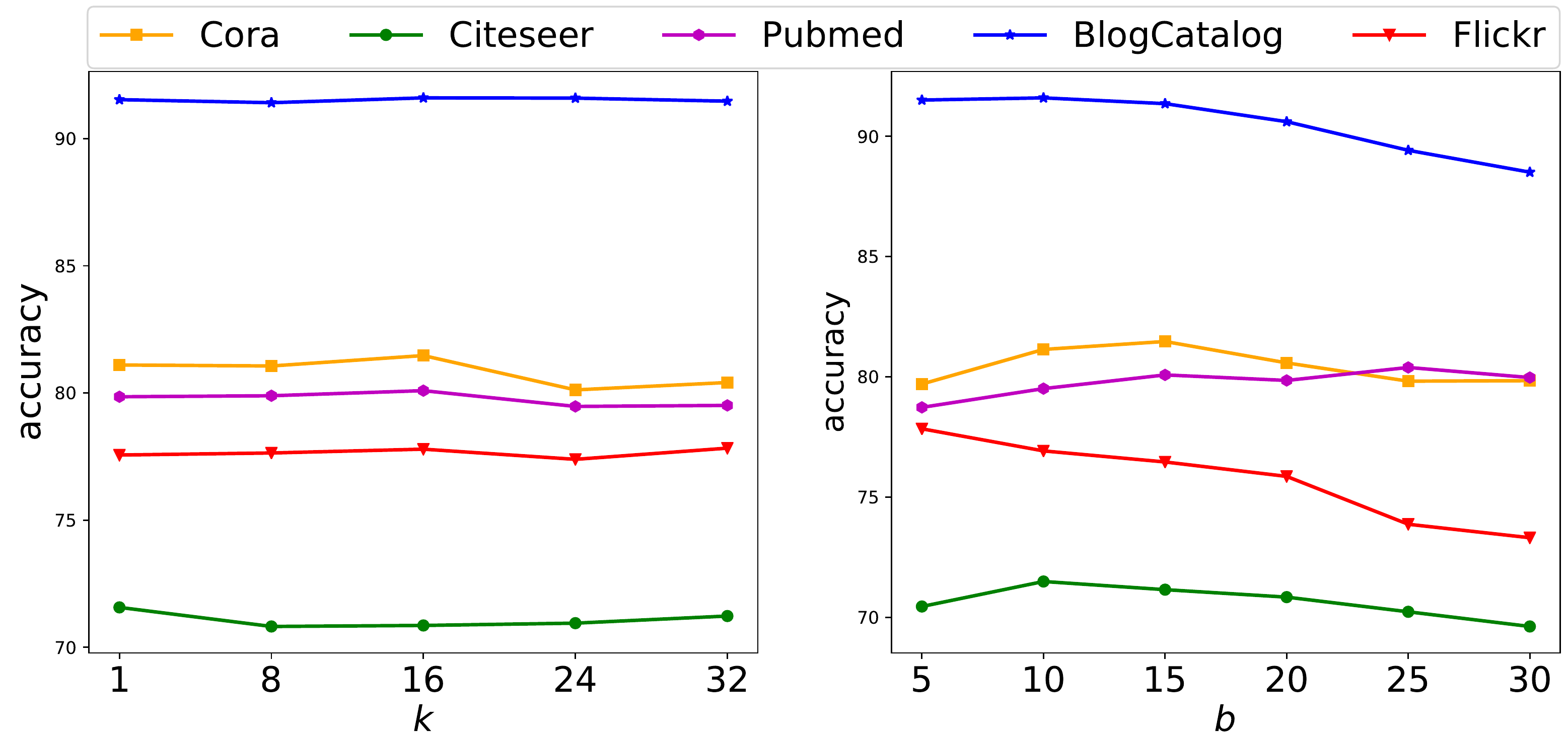}
\caption{Performances of UCMF with varying hyper-parameters: (1) the left part shows the impact of the negative sampling number $k$; (2) the right part shows the impact of the balance parameter $b$ between structure loss and classification loss.}
\label{fig:params}
\end{figure}

Experimental results are illustrated in Tab. \ref{tab:comparison1}-\ref{tab:comparison4} and Fig. \ref{fig:params}.

\begin{table}[t]
\centering
\scriptsize
\begin{tabular}{c|ccccc}
\hline
compared model & Cora  & Citeseer & Pubmed & BlogCatalog & Flickr \\
\hline
Planetoid & 75.7  & 64.3  & 77.2  & 84.7  & 70.9 \\
GCN   & \textbf{81.2} & 70.3  & 79.0    & 65.2  & 62.8 \\
fastGCN & 78.8  & 68.8  & 77.4  & 64.2  & 61.6 \\
SGC   & \textbf{81.0} & \textbf{71.9} & 78.9  & 58.8  & 37.2 \\
UCMF  & \textbf{81.4} & \textbf{71.5} & \textbf{80.4}$^*$ & \textbf{91.6}$^*$ & \textbf{77.8}$^*$ \\
\hline
\end{tabular}%
\caption{Performance comparison on semi-supervised node classification in terms of classification accuracy (\%) utilizing node features. The larger the values, the better the performances. Competitive performances on each dataset are highlighted. $*$ denotes statistically significant improvement measured by t-test with p-value $<0.05$.}
\label{tab:comparison2}
\end{table}

\textbf{The Roles of Unitization and Co-training (RQ1)}

Tab. \ref{tab:comparison1} illustrates the results of performance comparison without utilizing node features.
It is clear that, both UCMF-C and UCMF-U outperforms the conventional MF.
Meanwhile, UCMF can achieve better performances comparing with both UCMF-C and UCMF-U.
Moreover, UCMF and GCN perform closely on some datasets, and UCMF achieves the best performances on all datasets.
It is also worth to notice that, when we do not have node features, GCN has no obvious advantages comparing with DeepWalk and node2vec.
In average, UCMF relatively outperforms UCMF-C, UCMF-U and GCN $36.6\%$, $26.2\%$ and $13.0\%$ respectively.
These observations show that unitization and co-training are two useful and essential components of UCMF.

\textbf{UCMF vs. GCN (RQ2)}

Performance comparison on semi-supervised node classification are shown in Tab. \ref{tab:comparison2}.
It is obvious that, GCN outperforms fastGCN on all five datasets. This indicates that the edge sampling method may lead to somewhat performance loss.
We can also observe that, UCMF, SGC and GCN achieve competitive performances on Cora, Citeseer and Pubmed. This shows that, the two simplified GCN models inherit the effectiveness of GCN.
Moreover, on BlogCatalog and Flickr, both GCN and SGC have poor performances.
According to the sparsity of experimental datasets mentioned above, BlogCatalog and Flickr are relatively dense compared with the other three datasets. That is to say, models based on graph convolution perform poor on dense graphs. This may cause by the over-smoothing of graph convolution as mentioned in \cite{li2018deeper,chen2020measuring}, and this becomes more serious on dense graphs. Meanwhile, UCMF can balance the smoothing of neighbours and the classification of labeled nodes through the co-training process.
It is also worth to notice that, on BlogCatalog and Flickr, GCN outperforms SGC. This may indicate that, SGC aggravates the degree of over-smoothing of GCN.
In summary, UCMF achieves the best performances on all datasets except Citeseer. On Citeseer, UCMF performs very closely to SGC. Meanwhile, on BlogCatalog and Flickr, the advantages of UCMF are extremely large.
In average, UCMF relatively outperforms GCN, fast-GCN and SGC $12.3\%$, $14.7\%$ and $22.7\%$ respectively.
These results illustrate the effectiveness of UCMF.

\textbf{Stability Analysis (RQ3)}

As shown in Fig. \ref{fig:params}, we investigate the effects of the two important hyper-parameters, i.e., the negative sampling number $k$ and the balance parameter $b$, on the performances of UCMF.
The flat lines in the left part of Fig. \ref{fig:params} demonstrate that $k$ has little effects on the performances of UCMF.
That is to say, UCMF is stable with different negative sampling numbers.
As shown in the right part of Fig. \ref{fig:params}, on sparse datasets, i.e. Cora, Citeseer and Pubmed, $b$ has little impact on the performances of UCMF.
Meanwhile, on dense datasets, i.e., BlogCatalog and Flickr, the performances of the UCMF model slightly decrease with the increasing of $b$.
The hyper-parameter $b$ balances the structure loss and the classification loss. The larger value of $b$ indicates that UCMF pays more attention to capturing structural information.
That is to say, when UCMF pays too much attention to graph structure, it faces performance loss.
Combining with the viewpoint in \cite{li2018deeper,chen2020measuring}, this may give a reason for the poor performances of GCN on dense datasets: the graph convolution operation is easily to be extremely over-smoothing for capturing structural information on dense datasets.
Moreover, though performances of the UCMF model decrease when $b$ is large on dense datasets, the performance loss is slight. Overall, the performances of UCMF are relatively stable with hyper-parameters.
According to the observations from Fig. \ref{fig:params}, for other experiments, we set $k=16$ on all datasets, $b=15$ on sparse datasets, and $b=5$ on dense datasets.

\textbf{Distributed UCMF vs. Distributed GCN (RQ4)}

Performance comparison under distributed settings is shown in Tab. \ref{tab:comparison3}.
Recalling the results in Tab. \ref{tab:comparison2}, it is clear that both Cluster-GCN and Random-GCN greatly suffer performance loss.
Via sampling high-order neighbours of each node, fastGCN stays the performances as in Tab. \ref{tab:comparison2}.
Meanwhile, inherit from the capacity of linear model and MF, SGC and UCMF are extremely flexible for distributed computing, and both achieve very similar performances as in Tab. \ref{tab:comparison2}.
However, the performances of SGC are still poor on BlogCatalog and Flickr, because of the over-smoothing problem.
In average, under distributed settings, UCMF relatively outperforms Cluster-GCN, fast-GCN and SGC $19.8\%$, $15.4\%$ and $23.3\%$ respectively.
Overall, UCMF enlarges its advantages under distributed settings. These results strongly demonstrate the flexibility of UCMF.

\textbf{Results on Community Detection (RQ5)}

Tab. \ref{tab:comparison4} illustrates performance comparison on the task of community detection.
It is clear that, our proposed UCMF-C model can achieve best performances on all datasets.
This show that, the unitization module in UCMF is very useful, for it can reduce the possibility of extreme values, and has chance to generate better node representations.
These observations clearly demonstrate that, UCMF can perform well on the task of community detection, which is as typical task of graph embedding.

\begin{table}[t]
\centering
\scriptsize
\begin{tabular}{c|ccccc}
\hline
compared model & Cora  & Citeseer & Pubmed & BlogCatalog & Flickr \\
\hline
Random-GCN & 71.7  & 61.9  & 71.8  & 55.2  & 53.9 \\
Cluster-GCN & 77.1  & 64.8  & 76.6  & 59.8  & 57.5 \\
fastGCN & 77.6  & 68.7  & 77.1  & 64.1  & 61.2 \\
SGC   & \textbf{80.8} & \textbf{71.5} & 78.6  & 58.4  & 36.9 \\
UCMF  & \textbf{81.2}$^*$ & \textbf{71.3} & \textbf{80.3}$^*$ & \textbf{91.5}$^*$ & \textbf{77.6}$^*$ \\
\hline
\end{tabular}%
\caption{Performance comparison on semi-supervised node classification in terms of classification accuracy (\%) under distributed computing settings. The larger the values, the better the performances. Competitive performances on each dataset are highlighted. $*$ denotes statistically significant improvement measured by t-test with p-value $<0.05$.}
\label{tab:comparison3}
\end{table}

\begin{table}[t]
\centering
\scriptsize
\begin{tabular}{c|ccccc}
\hline
compared model & Cora  & Citeseer & Pubmed & BlogCatalog & Flickr \\
\hline
MF    & 25.5  & 27.5  & 26.6  & 76.5  & 50.1 \\
LINE  & 21.2  & 29.9  & 33.3  & 81.5  & 55.9 \\
DeepWalk & 26.3  & 20.2  & 21.6  & 76.2  & 49.6 \\
node2vec & 25.8  & 19.6  & 20.5  & 74.8  & \textbf{48.5} \\
UCMF-C & \textbf{12.3}$^*$ & \textbf{7.4}$^*$ & \textbf{14.3}$^*$ & \textbf{71.4}$^*$  & \textbf{48.1}$^*$ \\
\hline
\end{tabular}%
\caption{Performance comparison on community detection in terms of conductance. The smaller the values, the better the performances. Competitive performances on each dataset are highlighted. $*$ denotes statistically significant improvement measured by t-test with p-value $<0.05$.}
\label{tab:comparison4}
\end{table}

\section{Conclusion}
In this paper, we simplify GCN as unitized and co-training matrix factorization, and the UCMF model is therefore proposed. We conduct thorough and empirical experiments, which strongly verify our analysis. The experimental results on semi-supervised node classification show that UCMF achieves similar or superior performances compared with GCN. We also observe that GCN performs poor on dense graphs, while UCMF has great performances. This may be caused by the over-smoothing of graph convolution on dense graphs, while UCMF can balance the smoothing of neighbours and the classification of labeled nodes via co-training. Moreover, due to the MF-based architecture, UCMF is exceedingly flexible and convenient to be applied to distributed computing for real-world applications, and significantly outperforms distributed GCN methods. Meanwhile, on the task of community detection, our proposed UCMF model can achieve better performance compared with several representative graph embedding models. Extensive experimental results clearly show that, we can use UCMF as an alternative to GCN for transductive graph representation learning in various real-world applications.

\bibliography{UCMF}

\begin{thebibliography}{42}
\providecommand{\natexlab}[1]{#1}
\providecommand{\url}[1]{\texttt{#1}}
\providecommand{\urlprefix}{URL }
\expandafter\ifx\csname urlstyle\endcsname\relax
  \providecommand{\doi}[1]{doi:\discretionary{}{}{}#1}\else
  \providecommand{\doi}{doi:\discretionary{}{}{}\begingroup
  \urlstyle{rm}\Url}\fi

\bibitem[{Bo et~al.(2020)Bo, Wang, Shi, Zhu, Lu, and Cui}]{bo2020structural}
Bo, D.; Wang, X.; Shi, C.; Zhu, M.; Lu, E.; and Cui, P. 2020.
\newblock Structural Deep Clustering Network.
\newblock In \emph{Proceedings of The Web Conference 2020}, 1400--1410.

\bibitem[{Bruna et~al.(2013)Bruna, Zaremba, Szlam, and
  LeCun}]{bruna2013spectral}
Bruna, J.; Zaremba, W.; Szlam, A.; and LeCun, Y. 2013.
\newblock Spectral networks and locally connected networks on graphs.
\newblock \emph{arXiv preprint arXiv:1312.6203} .

\bibitem[{Chen et~al.(2020{\natexlab{a}})Chen, Lin, Li, Li, Zhou, and
  Sun}]{chen2020measuring}
Chen, D.; Lin, Y.; Li, W.; Li, P.; Zhou, J.; and Sun, X. 2020{\natexlab{a}}.
\newblock Measuring and Relieving the Over-smoothing Problem for Graph Neural
  Networks from the Topological View.
\newblock In \emph{AAAI Conference on Artificial Intelligence}.

\bibitem[{Chen, Ma, and Xiao(2018)}]{chen2018fastgcn}
Chen, J.; Ma, T.; and Xiao, C. 2018.
\newblock Fastgcn: fast learning with graph convolutional networks via
  importance sampling.
\newblock \emph{arXiv preprint arXiv:1801.10247} .

\bibitem[{Chen, Zhu, and Song(2018)}]{chen2018stochastic}
Chen, J.; Zhu, J.; and Song, L. 2018.
\newblock Stochastic training of graph convolutional networks with variance
  reduction.
\newblock In \emph{Neural Information Processing Systems}.

\bibitem[{Chen et~al.(2020{\natexlab{b}})Chen, Wei, Huang, Ding, and
  Li}]{Chen2020Simple}
Chen, M.; Wei, Z.; Huang, Z.; Ding, B.; and Li, Y. 2020{\natexlab{b}}.
\newblock Simple and Deep Graph Convolutional Networks.
\newblock In \emph{ICML}.

\bibitem[{Chiang et~al.(2019)Chiang, Liu, Si, Li, Bengio, and
  Hsieh}]{chiang2019cluster}
Chiang, W.-L.; Liu, X.; Si, S.; Li, Y.; Bengio, S.; and Hsieh, C.-J. 2019.
\newblock Cluster-GCN: An Efficient Algorithm for Training Deep and Large Graph
  Convolutional Networks.
\newblock In \emph{ACM SIGKDD Conference on Knowledge Discovery and Data
  Mining}.

\bibitem[{Du et~al.(2018)Du, Wang, Song, Lu, and Wang}]{du2018dynamic}
Du, L.; Wang, Y.; Song, G.; Lu, Z.; and Wang, J. 2018.
\newblock Dynamic Network Embedding: An Extended Approach for Skip-gram based
  Network Embedding.
\newblock In \emph{Proceedings of the Twenty-Seventh International Joint
  Conference on Artificial Intelligence}, 2086--2092.

\bibitem[{Gemulla et~al.(2011)Gemulla, Nijkamp, Haas, and
  Sismanis}]{gemulla2011large}
Gemulla, R.; Nijkamp, E.; Haas, P.~J.; and Sismanis, Y. 2011.
\newblock Large-scale matrix factorization with distributed stochastic gradient
  descent.
\newblock In \emph{ACM SIGKDD international conference on Knowledge discovery
  and data mining}, 69--77. ACM.

\bibitem[{Grover and Leskovec(2016)}]{grover2016node2vec}
Grover, A.; and Leskovec, J. 2016.
\newblock node2vec: Scalable feature learning for networks.
\newblock In \emph{ACM SIGKDD international conference on Knowledge discovery
  and data mining}, 855--864.

\bibitem[{Hamilton, Ying, and Leskovec(2017)}]{hamilton2017inductive}
Hamilton, W.; Ying, Z.; and Leskovec, J. 2017.
\newblock Inductive representation learning on large graphs.
\newblock In \emph{Advances in Neural Information Processing Systems},
  1024--1034.

\bibitem[{Hassani and Khasahmadi(2020)}]{hassani2020contrastive}
Hassani, K.; and Khasahmadi, A.~H. 2020.
\newblock Contrastive Multi-View Representation Learning on Graphs.
\newblock \emph{arXiv preprint arXiv:2006.05582} .

\bibitem[{He et~al.(2016)He, Zhang, Ren, and Sun}]{he2016deep}
He, K.; Zhang, X.; Ren, S.; and Sun, J. 2016.
\newblock Deep residual learning for image recognition.
\newblock In \emph{Proceedings of the IEEE conference on computer vision and
  pattern recognition}, 770--778.

\bibitem[{Hu et~al.(2019)Hu, Zhu, Wu, Wang, and Tan}]{hu2019hierarchical}
Hu, F.; Zhu, Y.; Wu, S.; Wang, L.; and Tan, T. 2019.
\newblock Hierarchical Graph Convolutional Networks for Semi-supervised Node
  Classification.
\newblock In \emph{Proceedings of the Twenty-Eighth International Joint
  Conference on Artificial Intelligence}, 10--16.

\bibitem[{Huang et~al.(2017)Huang, Liu, Van Der~Maaten, and
  Weinberger}]{huang2017densely}
Huang, G.; Liu, Z.; Van Der~Maaten, L.; and Weinberger, K.~Q. 2017.
\newblock Densely connected convolutional networks.
\newblock In \emph{Proceedings of the IEEE conference on computer vision and
  pattern recognition}, 4700--4708.

\bibitem[{Huang, Li, and Hu(2017)}]{huang2017label}
Huang, X.; Li, J.; and Hu, X. 2017.
\newblock Label informed attributed network embedding.
\newblock In \emph{Proceedings of the Tenth ACM International Conference on Web
  Search and Data Mining}, 731--739. ACM.

\bibitem[{Karypis and Kumar(1998)}]{karypis1998fast}
Karypis, G.; and Kumar, V. 1998.
\newblock A fast and high quality multilevel scheme for partitioning irregular
  graphs.
\newblock \emph{SIAM Journal on scientific Computing} 20(1): 359--392.

\bibitem[{Kipf and Welling(2017)}]{kipf2016semi}
Kipf, T.~N.; and Welling, M. 2017.
\newblock Semi-supervised classification with graph convolutional networks.
\newblock In \emph{International Conference on Learning Representations}.

\bibitem[{Levy and Goldberg(2014)}]{levy2014neural}
Levy, O.; and Goldberg, Y. 2014.
\newblock Neural word embedding as implicit matrix factorization.
\newblock In \emph{Advances in Neural Information Processing Systems},
  2177--2185.

\bibitem[{Li et~al.(2019)Li, Muller, Thabet, and Ghanem}]{li2019deepgcns}
Li, G.; Muller, M.; Thabet, A.; and Ghanem, B. 2019.
\newblock Deepgcns: Can gcns go as deep as cnns?
\newblock In \emph{Proceedings of the IEEE International Conference on Computer
  Vision}, 9267--9276.

\bibitem[{Li et~al.(2014)Li, Andersen, Park, Smola, Ahmed, Josifovski, Long,
  Shekita, and Su}]{li2014scaling}
Li, M.; Andersen, D.~G.; Park, J.~W.; Smola, A.~J.; Ahmed, A.; Josifovski, V.;
  Long, J.; Shekita, E.~J.; and Su, B.-Y. 2014.
\newblock Scaling distributed machine learning with the parameter server.
\newblock In \emph{11th $\{$USENIX$\}$ Symposium on Operating Systems Design
  and Implementation ($\{$OSDI$\}$ 14)}, 583--598.

\bibitem[{Li, Han, and Wu(2018)}]{li2018deeper}
Li, Q.; Han, Z.; and Wu, X.-M. 2018.
\newblock Deeper insights into graph convolutional networks for semi-supervised
  learning.
\newblock In \emph{Thirty-Second AAAI Conference on Artificial Intelligence}.

\bibitem[{Liu, Wu, and Wang(2015{\natexlab{a}})}]{liu2015collaborative}
Liu, Q.; Wu, S.; and Wang, L. 2015{\natexlab{a}}.
\newblock Collaborative prediction for multi-entity interaction with
  hierarchical representation.
\newblock In \emph{Proceedings of the 24th ACM International on Conference on
  Information and Knowledge Management}, 613--622. ACM.

\bibitem[{Liu, Wu, and Wang(2015{\natexlab{b}})}]{liu2015cot}
Liu, Q.; Wu, S.; and Wang, L. 2015{\natexlab{b}}.
\newblock Cot: Contextual operating tensor for context-aware recommender
  systems.
\newblock In \emph{Twenty-Ninth AAAI Conference on Artificial Intelligence}.

\bibitem[{Perozzi, Al-Rfou, and Skiena(2014)}]{perozzi2014deepwalk}
Perozzi, B.; Al-Rfou, R.; and Skiena, S. 2014.
\newblock Deepwalk: Online learning of social representations.
\newblock In \emph{ACM SIGKDD international conference on Knowledge discovery
  and data mining}, 701--710.

\bibitem[{Qiu et~al.(2018)Qiu, Dong, Ma, Li, Wang, and Tang}]{qiu2018network}
Qiu, J.; Dong, Y.; Ma, H.; Li, J.; Wang, K.; and Tang, J. 2018.
\newblock Network embedding as matrix factorization: Unifying deepwalk, line,
  pte, and node2vec.
\newblock In \emph{Proceedings of the Eleventh ACM International Conference on
  Web Search and Data Mining}, 459--467.

\bibitem[{Rendle et~al.(2011)Rendle, Gantner, Freudenthaler, and
  Schmidt-Thieme}]{rendle2011fast}
Rendle, S.; Gantner, Z.; Freudenthaler, C.; and Schmidt-Thieme, L. 2011.
\newblock Fast context-aware recommendations with factorization machines.
\newblock In \emph{Proceedings of the 34th international ACM SIGIR conference
  on Research and development in Information Retrieval}, 635--644. ACM.

\bibitem[{Sen et~al.(2008)Sen, Namata, Bilgic, Getoor, Gallagher, and
  Eliassirad}]{sen2008collective}
Sen, P.; Namata, G.; Bilgic, M.; Getoor, L.; Gallagher, B.; and Eliassirad, T.
  2008.
\newblock Collective Classification in Network Data.
\newblock \emph{Ai Magazine} 29(3): 93--106.

\bibitem[{Tang et~al.(2015)Tang, Qu, Wang, Zhang, Yan, and Mei}]{tang2015line}
Tang, J.; Qu, M.; Wang, M.; Zhang, M.; Yan, J.; and Mei, Q. 2015.
\newblock Line: Large-scale information network embedding.
\newblock In \emph{Proceedings of the 24th International Conference on World
  Wide Web}, 1067--1077.

\bibitem[{Veli{\v{c}}kovi{\'c} et~al.(2017)Veli{\v{c}}kovi{\'c}, Cucurull,
  Casanova, Romero, Lio, and Bengio}]{velivckovic2017graph}
Veli{\v{c}}kovi{\'c}, P.; Cucurull, G.; Casanova, A.; Romero, A.; Lio, P.; and
  Bengio, Y. 2017.
\newblock Graph attention networks.
\newblock In \emph{International Conference on Learning Representations}.

\bibitem[{Wang et~al.(2019)Wang, He, Wang, Feng, and Chua}]{wang2019neural}
Wang, X.; He, X.; Wang, M.; Feng, F.; and Chua, T.-S. 2019.
\newblock Neural graph collaborative filtering.
\newblock In \emph{Proceedings of the 42nd international ACM SIGIR conference
  on Research and development in Information Retrieval}, 165--174.

\bibitem[{Weston et~al.(2012)Weston, Ratle, Mobahi, and
  Collobert}]{Weston2012Deep}
Weston, J.; Ratle, F.; Mobahi, H.; and Collobert, R. 2012.
\newblock Deep Learning via Semi-supervised Embedding.
\newblock In \emph{International Conference on Machine Learning}.

\bibitem[{Wu et~al.(2019)Wu, Zhang, Souza~Jr, Fifty, Yu, and
  Weinberger}]{wu2019simplifying}
Wu, F.; Zhang, T.; Souza~Jr, A. H.~d.; Fifty, C.; Yu, T.; and Weinberger, K.~Q.
  2019.
\newblock Simplifying graph convolutional networks.
\newblock \emph{arXiv preprint arXiv:1902.07153} .

\bibitem[{Wu, He, and Xu(2019)}]{wu2019net}
Wu, J.; He, J.; and Xu, J. 2019.
\newblock Demo-net: Degree-specific graph neural networks for node and graph
  classification.
\newblock In \emph{Proceedings of the 25th ACM SIGKDD International Conference
  on Knowledge Discovery and Data Mining. ACM}.

\bibitem[{Wu et~al.(2016)Wu, Liu, Wang, and Tan}]{wu2016contextual}
Wu, S.; Liu, Q.; Wang, L.; and Tan, T. 2016.
\newblock Contextual operation for recommender systems.
\newblock \emph{IEEE Transactions on Knowledge and Data Engineering} 28(8):
  2000--2012.

\bibitem[{Xu et~al.(2018)Xu, Li, Tian, Sonobe, Kawarabayashi, and
  Jegelka}]{xu2018representation}
Xu, K.; Li, C.; Tian, Y.; Sonobe, T.; Kawarabayashi, K.-i.; and Jegelka, S.
  2018.
\newblock Representation learning on graphs with jumping knowledge networks.
\newblock \emph{arXiv preprint arXiv:1806.03536} .

\bibitem[{Yang, Cohen, and Salakhutdinov(2016)}]{Yang2016Revisiting}
Yang, Z.; Cohen, W.; and Salakhutdinov, R. 2016.
\newblock Revisiting Semi-Supervised Learning with Graph Embeddings.
\newblock In \emph{International conference on machine learning}.

\bibitem[{Ying et~al.(2018)Ying, He, Chen, Eksombatchai, Hamilton, and
  Leskovec}]{Ying2018Graph}
Ying, R.; He, R.; Chen, K.; Eksombatchai, P.; Hamilton, W.~L.; and Leskovec, J.
  2018.
\newblock Graph Convolutional Neural Networks for Web-Scale Recommender
  Systems.
\newblock In \emph{ACM SIGKDD International Conference on Knowledge Discovery
  \& Data Mining}.

\bibitem[{Yu et~al.(2020)Yu, Zhu, Liu, Wu, Wang, and Tan}]{yu2020TAGNN}
Yu, F.; Zhu, Y.; Liu, Q.; Wu, S.; Wang, L.; and Tan, T. 2020.
\newblock TAGNN: Target Attentive Graph Neural Networks for Session-based
  Recommendation.
\newblock In \emph{Proceedings of the 43th international ACM SIGIR conference
  on Research and development in Information Retrieval}. ACM.

\bibitem[{Yu et~al.(2014)Yu, Hsieh, Si, and Dhillon}]{yu2014parallel}
Yu, H.-F.; Hsieh, C.-J.; Si, S.; and Dhillon, I.~S. 2014.
\newblock Parallel matrix factorization for recommender systems.
\newblock \emph{Knowledge and Information Systems} 41(3): 793--819.

\bibitem[{Zhang and Chen(2018)}]{Zhang2018Link}
Zhang, M.; and Chen, Y. 2018.
\newblock Link Prediction Based on Graph Neural Networks.
\newblock In \emph{Neural Information Processing Systems}.

\bibitem[{Zhu et~al.(2020)Zhu, Xu, Yu, Liu, Wu, and Wang}]{zhu2020deep}
Zhu, Y.; Xu, Y.; Yu, F.; Liu, Q.; Wu, S.; and Wang, L. 2020.
\newblock Deep Graph Contrastive Representation Learning.
\newblock \emph{arXiv preprint arXiv:2006.04131} .

\end{thebibliography}
 
\end{document}